\documentclass[preprint2]{aastex}
 
\usepackage{epsfig}
\usepackage{graphicx}

\def\gx{GX~339$-$4}

\def\grs{GRS~1915$+$105}

\def\1e{1E~1740.7$-$2942}

\def\xte{XTE~J1550$-$564}

\def\1h{H~1743$-$322}
\def\4u{4U~1755$-$33}

\shorttitle{X-ray jets in \1h}
\shortauthors{Corbel et al.}

\begin{document}

\title{Discovery of X-ray Jets in the Microquasar \1h.}

\author{S. Corbel\altaffilmark{1}, P. Kaaret\altaffilmark{2}, R.P.
Fender\altaffilmark{3}, A.K. Tzioumis\altaffilmark{4}, J.A.
Tomsick\altaffilmark{5}, J.A. Orosz\altaffilmark{6}}
\altaffiltext{1}{AIM - Unit\'e Mixte de Recherche CEA - CNRS - Universit\'e Paris VII  - UMR 7158,
CEA Saclay, Service d'Astrophysique, F-91191 Gif sur Yvette, France.}
\altaffiltext{2}{Department of Physics and Astronomy, University of
Iowa, Iowa City, IA 52242 USA }
\altaffiltext{3}{ School of Physics and Astronomy, University of
Southampton, Highfield, Southampton, SO17 1BJ, England}
\altaffiltext{4}{Australia Telescope National Facility, CSIRO, P.O. Box
76, Epping NSW 1710, Australia. }
\altaffiltext{5}{Center for Astrophysics and Space Sciences, University
of California  at San Diego, MS 0424, La Jolla, CA92093, USA.}
\altaffiltext{6}{ Department of Astronomy, San Diego State University,
5500 Campanile Drive, PA 210, San Diego,  CA 92182-1221  USA}

\begin{abstract}

We report on the formation and evolution of two large-scale,
synchrotron-emitting jets from the black hole candidate \1h\ following
its reactivation in 2003.  In November 2003 after the end of its 2003
outburst, we noticed, in observations with the Australia Telescope
Compact Array, the presence of a new and variable radio source about
4.6\arcsec\ to the East of \1h, that was later found to move away from
\1h.  In February 2004, we detected a radio source to the
West of \1h, symmetrically placed relative to the Eastern jet.  In
2004, follow-up X-ray observations with {\em Chandra} led to the
discovery of X-ray emission associated with the two radio sources. 
This likely indicates that we are witnessing the interaction of
relativistic jets from \1h\ with the interstellar medium causing
in-situ particle acceleration.  
The spectral energy distribution of the jets during the decay phase
is  consistent with a classical synchrotron spectrum of a single
electron distribution from radio up to X-rays, implying the production
of very  high energy ($>$ 10 TeV) particles in those jets.  
We discuss the jet kinematics, highlighting the presence of  a significantly
relativistic flow in \1h\ almost a year after the ejection event. 

\end{abstract}


\section{Introduction}

Relativistic jets are now believed to be a common occurrence in black
hole X-ray binaries (e.g. Corbel 2004, Fender 2004). The optically thin
synchrotron spectra of the discrete ejection events (the so-called
superluminal jets) imply that the emission becomes much fainter at
higher frequencies.  For that reason, such events were only observed at
radio frequencies. However, Sams et al.\ (1996) reported the detection
of extended infrared emission in \grs\ after a massive ejection event
that could be of synchrotron nature.  The most extreme case has been
the detection of moving and decelerating X-ray (and radio) relativistic
jets in the microquasar \xte\ a few years after the ejection event and
on over large distances (Corbel et al.\ 2002; Tomsick et al.\ 2003a;
Kaaret et al.\ 2003).  The detection of optically thin synchrotron
X-ray emission from discrete ejection events implies in-situ particle
acceleration up to several TeV.  This acceleration may be caused by
interaction of the jets with the interstellar medium (ISM).  

It now appears that relativistic jets emit throughout the entire 
electromagnetic spectrum.  But are the jets of \xte\ unique due to
special physical conditions or should we expect to observe similar
X-ray jets in other X-ray Novae (XNe) ?  If jets containing TeV particles are common
amongst XNe, then there may be interesting consequences.  The high
energy particles could produce a distinctive signature in the
cosmic-ray flux (Heinz \& Sunyaev 2002) or produce neutrinos if the
jets contain protons (Distefano et al. 2002).  Furthermore, the study
of X-ray jets in microquasars could add an additional bridge between
the physics of jets in microquasars and those from supermassive black
holes.  Interestingly, a connection with Gamma-Ray Burst has also been
drawn by Wang, Dai \& Lu (2003), who showed that the evolution of the
eastern  X-ray jet of \xte\ was consistent with the emission from
adiabatically  expanding ejecta heated by a reverse shock following its
interaction with the ISM.  The sensitivity of current X-ray missions
(in particular {\em Chandra} and {\em XMM-Newton}) is sufficient to
enable new discoveries in this emerging field and therefore provides
further clues regarding the physical nature of these high energy
phenomena.

\1h\  was discovered with Ariel 5  in August 1977 (Kaluzienski \& Holt
1977) and has been precisely localized  by HEAO-1 (Doxsey et al. 1977).
It was originally proposed to be a black hole  candidate (BHC) by White
\& Marshall (1983). \1h\ has probably been active  several times during
the past decades with activity observed in 1984 by   EXOSAT (Reynolds
et al. 1999) and in 1996 by TTM (Emelyanov et al. 2000).   In March
2003, INTEGRAL detected new activity from IGR~J17464$-$3213 
(Revnivtsev et al. 2003) that was later found to correspond to \1h. 
After its reactivation in 2003, a radio counterpart was found with the
VLA  by Rupen, Mioduszewski \& Dhawan (2003a)  and a bright radio flare
(likely associated with a massive ejection event) was observed on 2003
April 8 (MJD 52738) by Rupen, Mioduszewski \& Dhawan (2003b). During
outburst, \1h\ went through several X-ray states with properties
typical of BHC.  Low and high frequency quasi-periodic oscillations
have been detected in \1h\ by {\em RXTE} (Homan et al. 2005; Remillard
et al. 2005). {\em Chandra} observations, during the 2003 outburst,
revealed the presence of narrow and variable absorption lines in the
X-ray spectrum of \1h,  possibly related to an ionized outflow (Miller
et al. 2005). The 2003 active phase (as observed by RXTE/ASM) ended
around late November 2003, but a new 4-month outburst started in July
2004 (Swank 2004) with a later reactivation at radio frequencies
(Rupen, Mioduszewski \& Dhawan 2004).

In this paper, we present an analysis of {\em Chandra} X-ray and ATCA
radio observations of \1h\ from November 2003 through June 2004 that
led to the discovery of large scale radio jets on each side of the BHC
\1h.  More importantly, we also report the discovery of X-ray emission
is very reminiscent of the large  scale decelerating relativistic jets
observed in \xte\ (Corbel et al. 2002). In \S2, we describe the
detection and  localization of the radio and X-ray sources.  In \S3, we
discuss the jet kinematics and the nature  of the jet emission
mechanism.

\section{Observations and sources detection}

\begin{table*}[ht]
\caption{\label{tab_logr}  Radio Observations of \1h\ and associated jets: position and angular separation. }
\begin{tabular}{lllccc}
\noalign{\smallskip}
\hline \hline
\noalign{\smallskip}
Date       & MJD       & Source       & \multicolumn {2}{c}{Position} & Angular \\
\cline{4-5}
\noalign{\smallskip}
           &           &              &    R.A.      & Decl.     & Separation \\
\hline \hline
\noalign{\smallskip}
2003/11/12 & 52955.86  &  Eastern Jet & 17 46 15.978 & -32 14 01.08 & 4.63 $\pm$ 0.30\arcsec \\
\hline
2003/11/30 & 52973.72  &  Eastern Jet & 17 46 15.995 & -32 14 00.15 & 4.91 $\pm$ 0.30\arcsec \\
\hline
2003/12/09 & 52983.45  &  Eastern Jet & 17 46 16.005 & -32 14 00.83 & 4.98 $\pm$ 0.30\arcsec \\
\hline 
2003/12/20 & 52994.49  &  Eastern Jet & 17 46 16.026 & -32 14 00.67 & 5.25 $\pm$ 0.30\arcsec \\
\hline
2004/02/13 & 53049.40  &  Eastern Jet & 17 46 16.056 & -32 14 00.40 & 5.65 $\pm$ 0.50\arcsec \\
           &           &  Western Jet & 17 46 15.274 & -32 14 00.78 & 4.30 $\pm$ 1.00\arcsec \\
\hline
2004/04/07 & 53103.22  &     \multicolumn {4}{c}{No detection} \\
\hline
2004/06/01 & 53158.15 &      \multicolumn {4}{c}{No detection} \\
\noalign{\smallskip}
\hline \hline
\end{tabular}
\vspace{0.3cm}

MJD: Modified Julian Date (observation midtime). The angular separation
is based on the {\em Chandra} position of the black hole \1h. 
\end{table*}

\begin{table*}[ht]
\caption{\label{tab_logr}  Radio Observations of \1h\ and associated jets: flux density and spectral index. }
\begin{minipage}{\linewidth}
\hspace{-1cm}
\scriptsize
\begin{center}
\begin{tabular}{lllccccc}
\noalign{\smallskip}
\hline \hline
\noalign{\smallskip}
      &       &        & \multicolumn{4}{c}{Flux density (mJy)} & \\
Date       & MJD       & Source       &     \multicolumn{4}{c}{ MHz}     &  Spectral index        \\ 
\cline{4-7}
\noalign{\smallskip}
           &           &              &   1384 & 2368 & 4800            & 8640  &  \\
\hline \hline
\noalign{\smallskip}
2003/11/12 & 52955.86  &  Eastern Jet &  ...   &  ... & 1.10 $\pm$ 0.10 & 0.48 $\pm$ 0.11 &  $-$1.41 $\pm$ 0.42  \\
\hline
2003/11/30 & 52973.72  &  Eastern Jet &  ...   &  ... & 1.55 $\pm$ 0.09 & 0.87 $\pm$ 0.09 &  $-$0.98 $\pm$ 0.20 \\
\hline
2003/12/09 & 52983.45  &  Eastern Jet &  ...   &  ... & 1.86 $\pm$ 0.09 & 0.84 $\pm$ 0.09 &  $-$1.35 $\pm$ 0.20  \\
\hline
2003/12/20 & 52994.49  &  Eastern Jet &  ...   &  ... & 2.37 $\pm$ 0.06 & 1.34 $\pm$ 0.07 &  $-$0.97 $\pm$ 0.10  \\
\hline
2004/02/13 & 53049.40  &  Eastern Jet &  0.62 $\pm$ 0.15 &  0.41 $\pm$ 0.15 & 0.33 $\pm$ 0.05 & $<$ 0.21  & $-$0.49 $\pm$ 0.22 \\
           &           &  Western Jet &  $<$ 0.45        &  $<$ 0.45        & 0.14 $\pm$ 0.05 & $<$ 0.21  & N.A. \\
\hline
2004/04/07 & 53103.22  & Eastern Jet &  ...   &  ... &  $<$ 0.18 & ... & N.A.    \\
\hline
2004/06/01 & 53158.15  & Eastern Jet &  $<$ 0.45  &  $<$ 0.30 &  $<$ 0.12 & ... & N.A.    \\
\noalign{\smallskip}
\hline \hline
\end{tabular}
\vspace{0.3cm}

MJD: Modified Julian Date (observation midtime). Radio upper limits are
given at the 3 sigma level.  Fluxes for the western jet are only given
for the day of its unique ATCA detection, otherwise the upper limits
can be deduced from the error bars (one sigma) or upper limits on the
flux densities of the eastern jet.

\end{center}

\end{minipage}

\end{table*}

\subsection{Radio observations}

Following the transition of \1h\ to the hard X-ray state (Tomsick \&
Kalemci 2003) on 2003 October 20  (MJD 52933), we initiated a series of
radio observations with the Australia Telescope Compact  Array (ATCA)
starting on 2003 November 12 (MJD 52955).  The ATCA synthesis telescope
is located in Narrabri, New South Wales, Australia and consists of an
east-west array with six 22~m antennas.   A total of seven radio
observations were carried out with the ATCA, mostly at 4800 MHz and
8640 MHz with occasional observations at 1384 MHz and  2496 MHz (see
Tables 1 and 2 for details).  The amplitude and band-pass calibrator
was PKS~1934$-$638, and the antenna's gain and phase calibration, as
well as the polarization leakage, were derived from regular
observations of the nearby (5.8\degr\ away) calibrator PMN~1729$-$373.
The editing, calibration, Fourier transformation, deconvolution, and
image analysis were performed using the MIRIAD software package (Sault
\& Killen 1998).

No radio emission is detected at the position of \1h\ in any of our
ATCA observations.  However, we notice the presence of a new radio
source (Fig. 1) about 4.6\arcsec\ to the East of \1h, immediately in the first
ATCA observation.  The following observations indicate that the new
source increased in flux (see Tables 1 and 2 and Fig. 2).  The position
of the source also changed over time, moving away from the BHC \1h.  We
discuss the motion of the source in a later section taking into account
the additional constraints from the {\em Chandra} observations.

The light-curve of this source, combining all our ATCA observations (at
4800 MHz and 8640 MHz), is presented in Figure 2. This source was
brightening from our first observations until the end of 2003.  The
single detection on 2004 February 13 (MJD 53049) at a much (almost  a
factor ten) fainter level points to a very fast decay. The maximum of
radio emission was likely in January 2004.  The rise of radio emission
at 4800 MHz is gradual, whereas there may be some variations during the
rise at 8640 MHz.  We fitted an exponential rise and a power-law rise
to the radio data (at both frequencies) and found that both fits are
equally acceptable.  The time-scales are consistent for both radio
frequencies.  The 1/{\it e}-folding time of the exponential rise  is
49.4 $\pm$ 4.8 days at 4800 MHz and 40.0 $\pm$ 6.8 days at 8640 MHz.
The index of the power-law rise is 4.9 $\pm$ 0.5  at 4800 MHz and 6.0
$\pm$ 1.0 at 8640 MHz.

Due to the fast decay of radio emission, we can not accurately
constrain its decay rate.  However, we can give some limits that
clearly indicate that the decay is faster than the rise.  Assuming a
maximum at the time when we detected the source at its brightest level
(on 2003 December 20), and taking the 4800 MHz detection during the
decay, we deduced an 1/{\it e}-folding time of the exponential decay of
27.9 $\pm$ 2.2 days or an index of  the power-law decay of 10.2 $\pm$
0.8.  We emphasize that these limits (an upper limit on the 1/{\it
e}-folding  time and a lower limit on the index of the power-law) are
quite conservative and that the true decay during the early part of
2004 is likely faster than indicated by these numbers.  The light-curve
at 8640 MHz is consistent with that at 4800 MHz, but very few data
points are available.

We searched for linearly polarized radio emission when the eastern
source was at its brightest level (i.e. on 20 December 2003). Linear
polarization at the level of 9.9 $\pm$ 2.9\%\ with a mean polarization
angle of 51.9 $\pm$ 8.3\degr\ is detected at 4800 MHz.  At 8640
MHz, we find an (3 sigma) upper limit of $\sim 21 \% $ on the polarization.  This
limit is weak because the 8640 MHz flux density is weaker than at 4800 MHz.

In addition to the new source to the East of \1h, we note that a radio
source is detected to the West of \1h\ (Figure 3), almost symmetrically
placed relative to the Eastern source, only on 2004 February 13 with a
flux density of 0.14 $\pm$ 0.05 mJy  at 4800 MHz. Despite its weak
flux  density, the {\em Chandra} data on the same day (Figs. 4 and 5)
confirm the  existence of an X-ray source at this position and
therefore strengthen the reality of the western radio source. 

As both of these radio sources appear to be moving away from the BHC, 
we conclude that they are related to the  \1h\ system.  The properties
are very reminiscent to the large scale relativistic radio and X-ray
jets of \xte\ (Corbel et al. 2002) after 2000.  It is therefore likely
that these evolving radio sources represent the action of previously
ejected plasma on the ISM.  As discussed below, the most likely
ejection date is around 2003 April 8 when a major radio flare was
observed by Rupen  et al.\ (2003).  We will now refer to these sources
as the eastern and western jets, based on their location with respect
to \1h.  We report in Tables 1 and 2 their positions (from radio
observations), angular separation from \1h, as well as their flux
densities.  As \1h\ is not detected at radio frequencies, we used its
{\em Chandra} position as defined in the next section.

\subsection{X-ray observations}

\subsubsection{Source detection}

Following the detection of the eastern moving radio source and its
strong similarities with the large scale  X-ray jets in \xte\ (Corbel
et al. 2002; Tomsick et al. 2003a; Kaaret et al. 2003), we triggered our
target of opportunity proposal for the {\em Chandra X-ray Observatory}
(Weisskopf et al. 2002) to search for an X-ray counterpart to the radio
jets.  The {\em Chandra} observations were significantly delayed
relative to the trigger because \1h\ passed near the Sun and was not
observable.  As indicated previously, the eastern jet was already
decaying when the {\em Chandra} observations were performed (see
Fig. 2 for their scheduling relative to the radio light-curve).

\1h\ was observed with {\em Chandra} on 2004 February 12 (MJD 53048),
March 24 (MJD 53089) and March 27 (MJD 53092) using the Advanced CCD
Imaging Spectrometer spectroscopic array (ACIS-S; Bautz et al. 1998).
In all observations, the target was placed on one of the 
back-illuminated ACIS Chips (S3) with the ACIS-S operated in imaging
mode.  For the first observation, only the S3 chip was read out and a
1/2 sub-array mode was used to limit pile-up.  For the latter two
observations, the source was known to be a lower flux state and the
full ACIS-S imaging mode array was used.

We produced 0.3--8 keV ACIS images using the ``level 2'' event lists
from the standard data processing  (ASCDS v7.2.1) using the {\em Chandra}
Interactive Analysis of Observations (CIAO) software package version
3.0.2 and Calibration Data Base (CALDB) version 2.26.  We constructed
light curves with all valid events on the  S3 chips to search for time
of high background.  Only weak background flares were found (and
removed) for  the February observation, otherwise the count rate
appears uniform.  The total useful exposure obtained was 17796~s on
February 12 (Obs.\ 1), 28363~s on March 24 (Obs.\ 2), and 40037~s
on March 27 (Obs.\ 3).

We searched for X-ray sources in each 0.3--8 keV image using {\it
wavdetect} (Freeman et al.\ 2002),  the wavelet-based source detection
routine in CIAO.  For all three {\em Chandra} observations, an X-ray
source is found at the location of \1h, the eastern jet, and western
jet (Figs 4 and 5).  All three sources appear aligned and therefore
provide further evidence for a connection with \1h.  

We check the {\em Chandra} {\it wavdetect} position of each source by
calculating the source's centroid using the 0.3-8 keV events from  a 4
$\times$ 4 pixels (2\arcsec $\times$ 2\arcsec) region centered on the
{it wavdetect} positions.  This region contains all events likely
related to the source along with a small number of background events.
The recalculated positions of \1h, the eastern and western jets were in
good agreement with the {\it wavdetect} positions; the differences were
less than $0\arcsec .10$.  The position angle of the eastern jet
relative to \1h\ is 89.0 $\pm$ 1.5 \degr, while the position angle of
the western jet is --91.7 $\pm$ 1.8 \degr.

\subsubsection{Source localization}

To obtain the best constraint on the position of each detected source
(especially  \1h, which is not detected in our radio images), we
registered the {\em Chandra } images with an infrared image from the 
2 Micron All-Sky Survey (2MASS) following the procedure described in
Tomsick et al.\ (2003b).  We use the two longest {\em Chandra} 
observations (\# 2 and 3) and inspected each observation separately. 
We restricted our search for X-ray and infrared sources within a 4
arcmin radius circle centered on the position of \1h\ as reported by 
{\it wavdetect}.

We cross-correlated the {\em Chandra } source positions with the 2MASS
sources in the field.   We find that 5 {\em Chandra } sources have
2MASS sources within the {\em Chandra } pointing uncertainty of
0\arcsec.6 for observation 2 and 9 for observation 3.  The accuracy of
the 2MASS source positions is 0\arcsec.2 (90\% confidence).  For these
sources, the angular  separation between the 2MASS and {\em Chandra }
positions ranges from 0\arcsec.07 to  0\arcsec.22 for observation 2 and
from  0\arcsec.06 to 0\arcsec.46 for observation 3. Given the surface
density of 2MASS sources, there is a 0.11\% probability that a match
with the largest separation is spurious  for observation 2, and 0.50\%
for observation 3. We used the full set of matches to register each
X-ray image.  The average 2MASS to {\em Chandra } differences are
$-0\arcsec .01 \pm 0\arcsec .11$ in R.A. and $-0\arcsec .05 \pm
0\arcsec .10$ in Decl.\  for observation 2  and $-0\arcsec .19 \pm
0\arcsec .10$ in R.A.\ and  and $-0\arcsec.09 \pm 0\arcsec.08$ for
observation 3.  We estimated the location of \1h\ as the average of the
positions measured in the two {\em Chandra } images after registration
to the 2MASS positions by performing the above indicated shifts. \1h\
is found to be located at R.A. = 17h 46m 15.s613, Decl.  = $-$32\degr\
14\arcmin\ 0\arcsec .95 with a total uncertainty of 0.\arcsec 10
(including the 0.2\arcsec\ systematics due to the registration to 2MASS
images and the statistical uncertainties on the {\em Chandra } {\it
wavdetect} position).  The difference in position of the eastern and
western jets, after the registering process, between the two {\em
Chandra } observations (only separated by 2.5 days) were less than
0.\arcsec 10 (i.e. within the total  uncertainty).  This precise {\em
Chandra } position of \1h\ is consistent within uncertainties (a
difference of 0.25\arcsec) with the recently refined VLA radio position
(Rupen, Mioduszewski, \& Dhawan 2004).  

We used the position of \1h\ to register {\em Chandra } observation \#
1 on the 2MASS frame, for which shifts of $-$0.25 $\pm$ 0.17 arcsecond
(in R.A.) and $-$0.32 $\pm$ 0.16 arcsecond (in Decl.) are needed.
These shifts are smaller than the {\em Chandra} absolute astrometric of
$0.6\arcsec$ at 90\% confidence.  We report in Table 3 the location
of  \1h , the western and eastern jets, as well as the angular
separation between  the jets and the BHC for all three {\em Chandra }
observations. We note that the angular  separation does not depend on
the image registration. 

\begin{table*}[ht]
\caption{\label{tab_log} {\em Chandra} Observations of \1h. }
\begin{minipage}{\linewidth}
\hspace{-1cm}
\scriptsize
\begin{center}
\begin{tabular}{cllclccccc}
\noalign{\smallskip}
\hline \hline
\noalign{\smallskip}
Obs. & Date & MJD & Exposure & Source & \multicolumn {3}{c}{Position} & Angular  \\
\cline{6-8}
\noalign{\smallskip}
\#   & 2004-     &     &  time (s)          &        &    R.A.  & Decl. & Uncertainty & Separation   \\
\hline \hline
\noalign{\smallskip}
1 & 02-12  & 53048.0 & 17796 &  \1h\ & 17 46 15.613 & -32 14 0.95 & 0.10\arcsec\ & N.A.  \\
 &    &       &       & Eastern Jet & 17 46 16.094 & -32 14 0.92 & 0.19\arcsec\ & 6.10 $\pm$ 0.20 \arcsec\ \\
 &    &       &       & Western Jet & 17 46 15.263 & -32 14 1.44 & 0.24\arcsec\ & 4.47 $\pm$ 0.30 \arcsec\ \\
\hline
\noalign{\smallskip}
2 & 03-24  & 53088.9 & 28363 & \1h\ & \multicolumn {3}{c}{See Obs. \# 1} & N.A.  \\
 &    &       &       & Eastern Jet & 17 46 16.142 & -32 14 0.76 & 0.17\arcsec\ & 6.72 $\pm$ 0.20 \arcsec\ \\
 &    &       &       & Western Jet & 17 46 15.243 & -32 14 1.05 & 0.16\arcsec\ & 4.70 $\pm$ 0.20 \arcsec\ \\
\hline
\noalign{\smallskip}
3 & 03-27  & 53091.5 & 40037 & \1h\ & \multicolumn {3}{c}{See Obs. \# 1} & N.A.  \\
 &    &       &       & Eastern Jet & 17 46 16.135 & -32 14 0.83 & 0.11\arcsec\ & 6.63 $\pm$ 0.20 \arcsec\ \\
 &    &       &       & Western Jet & 17 46 15.243 & -32 14 1.09 & 0.15\arcsec\ & 4.70 $\pm$ 0.20 \arcsec\ \\
\noalign{\smallskip}
\hline \hline
\end{tabular}
\vspace{0.3cm}

MJD: Modified Julian Date (exposure midtime). The angular separation between the jets and H~1743$-$322 
is based on the {\em Chandra} position of the black hole H~1743$-$322
as estimated in observation \# 2 and 3 during the 2MASS/{\em Chandra} registration process. The uncertainties on the 
position include systematics due to the registration process and the statistical uncertainties on the {\it wavdetect} 
position. However, the angular separation does not depend on the image registration.  

\end{center}

\end{minipage}

\end{table*}

\subsubsection{Flux and energy spectra}

We extracted an energy spectrum in the 0.3--8 keV energy range for the
eastern and western jets in all three {\em Chandra} observations using
CIAO v3.0.2 tools and we fitted these spectra using XSPEC v11.3.0.   We
used a circular source extraction region with a radius of 2.3\arcsec,
resp.\  1.2\arcsec, and we extracted background spectra from an annulus
with an inner radius of 13\arcsec\ and an outer radius of 22\arcsec,
resp. \ 19.5\arcsec\ for the eastern and western jets respectively.
These regions were centered on the jet positions  as given by {\it
wavdetect}. Due to the low numbers of source counts (at maximum of 24
counts),  we used the W statistic for fitting (Wachter, Leach \& Kellog 1979; 
Arnaud, in prep.) the un-binned spectra, this is adapted from the Cash 
statistic (Cash 1979) and works with background subtracted spectra
and low count rate. 

These spectra are adequately fitted with a power-law model including
interstellar absorption.  We fixed the  equivalent hydrogen absorption
column density, N$_\mathrm{H}$, to the constant value measured in {\em
Chandra} observations of \1h\ during its 2003 outburst (Miller et al.
2005), i.e. 2.3 $\times$ 10$^{22}$ cm$^{-2}$.  For the eastern jet, the
best-fit photon index is 1.67 $\pm$ 0.90, 1.78 $\pm$ 1.10 and 2.03
$\pm$ 0.90 for  {\em Chandra} observations \# 1, 2 and 3 respectively.
Refitting these three datasets simultaneously (allowing the normalization 
to vary) led to a photon index of 1.83 $\pm$
0.54 (with 90\% confidence errors).  We did the same for the western
jet, giving photon indexes of 2.5 $\pm$ 2.5, 2.2 $\pm$ 1.0 and 1.6 $\pm$
1.0 for {\em Chandra} observations \# 1, 2 and 3 respectively and a
photon index of 1.9 $\pm$ 0.8 (with 90\%  confidence errors) for the
combined spectrum.

We fixed the power-law photon index (see section 3.2.2) to a value of
1.6 to obtain measurements of the X-ray flux.  This value is extracted
from the fit to the radio/X-ray spectral energy distribution. We
measure 0.3--8 keV absorbed fluxes for the eastern jet of (21.3 $\pm$
4.9) $\times$ 10$^{-15}$,  (10.0 $\pm$ 2.6) $\times$ 10$^{-15}$,  and
(9.4 $\pm$ 2.3) $\times$ 10$^{-15}$ ergs s$^{-1}$ cm$^{-2}$ for
observations 1, 2, and 3, respectively. Concerning the western jet, the
0.3--8 keV absorbed fluxes are (3.0 $\pm$ 2.1) $\times$ 10$^{-15}$, 
(6.4 $\pm$ 2.6) $\times$ 10$^{-15}$, and (6.2  $\pm$ 1.8) $\times$
10$^{-15}$ ergs s$^{-1}$ cm$^{-2}$ for  observations 1, 2, and 3,
respectively.  The unabsorbed fluxes, in the same band, are 2.0 higher
than the absorbed values quoted above. The quoted errors are based on
the numbers of source and background counts and Poisson statistics.

Figure 6 shows the time variation of the 0.3--8 keV unabsorbed flux of
the  eastern and western jets.  The X-ray emission of the eastern jet
is decaying, whereas the emission from the  western jet is consistent
with a slightly rising source (or even constant). We fitted an
exponential  or a power-law decay to the eastern jet and we found that
both fits are equally acceptable. The 1/{\it e}-folding  time of the
exponential decay is 53.5 $\pm$ 19.6 days, whereas the index of the
power-law decay is 6.2  $\pm$ 2.3. The constraints are rather weak for
the western jet, the 1/{\it e}-folding time of the  exponential rise is
58.2 $\pm$ 46.3 days and the index of the power-law rise is 5.7 $\pm$
4.5. 

We note that the western jet was likely detected at 4.8 GHz during the
first {\em Chandra} observation (Figure 3), but unfortunately no radio
observation was possible during the time of the second and third {\em
Chandra} observations. However, we did perform a ATCA observation about
2 weeks after the third {\em Chandra}  observations but failed to
detect radio emission from the western jet (with similar sensitivity to
its detection in February, see Table 2). This indicates that the peak
of emission from the western jet was likely sometime between February
and the beginning of April 2004, i.e. roughly 2 or 3 months after the
peak of emission from the eastern jet. At this peak, the X-ray (0.3 --
8 keV band) luminosity (for a distance of 8 kpc)  of the eastern jet
was of the order of more than 3 $\times$ 10$^{32}$ ergs s$^{-1}$, i.e.
a level consistent  with the X-ray emission of some BHCs while in their
quiescent state (e.g. Garcia et al. 2001).  Therefore,  if the 
production of X-ray jets is more common than previously thought as we
suggest here, some of the reported measurements of the quiescent X-ray
emission of BHC may be contaminated by jet emission, particularly for
observations made using instruments with worse angular resolution than
{\em Chandra}.

\subsubsection{Source morphology}
\label{xmorph}

The western jet of \xte\ was clearly extended with a leading peak and a
trailing tail extended  up to 5\arcsec\ toward the black hole (Kaaret
et al. 2003).  To study the morphology of the X-ray jets in \1h, we
decomposed the images along axes parallel and perpendicular to the jet
axis defined as a line extending through the black hole candidate and
the jets. Following the procedure described in Kaaret et al.\ (2003),
we calculated the displacement of each X-ray event parallel and
perpendicular to this axis.  All photons with energies in the range
0.3--8~keV and within $2\arcsec$ of the jet axis in the perpendicular
direction are included.  For each observation, we compared the
morphology of the black hole candidate  \1h, assumed to be point-like,
to each of the detected jets, except for observation 1 in which the
Western jet had too few counts for a meaningful comparison.

We carried out the morphology comparison using a Kolmogorov-Smirnov
(KS) test to permit comparison of unbinned position data.  We shifted
the jet positions to match the black hole candidate position using the
positions derived from {\it wavdetect}. We compared the morphology for
events lying with $2\arcsec$ along the jet axis of each source
position.  We find no significant evidence for spatial extent along or
perpendicular to the jet axis for any jet in any observation.  We also
combined the data from observations 2 and 3  (Fig. 7) and still did not
find any significant evidence for spatial extent of either jet.  The
KS-test probabilities that the black hole and jet samples are drawn
from the same parent distribution range from 0.11 to 0.96.  To place an
upper bound on the size of the Western jet along the jet axis, we
calculated the standard deviation of the displacements of events along
the jet axis from the {\it wavdetect} position of the jet. We find a
value of $0.61 \pm 0.04\arcsec $.  This is inconsistent with the
standard deviation calculated for the black hole candidate 
(0.48 $\pm$ 0.02\arcsec\ ) by only $2.7\sigma$, so, again, there is no strong evidence
for spatial extent and the value should be taken as an upper limit on
the jet size.

In order to increase the statistics for the comparison source, we decided
to compare the X-ray jet profiles to the profiles for the source PG~1634$+$706 
(observation ID 1269), which is used to calibrate the ACIS point-spread function. 
Again, the KS test does not indicate significantly that the BHC \1h\ or its 
associated X-ray jets are extended along or perpendicular to the jet axis. 

\section{Discussion}

\subsection{Proper motion and jet kinematics}

The positions of the eastern and western jets change between the
various radio and X-ray observations with the jets are moving away from
the BHC \1h.  In order to quantify this motion, we have calculated the
angular separation between \1h\ and each jet.  For all {\em Chandra}
and ATCA observations, we used the position of \1h\ as determined in
the previous section.  Figure 8 shows the angular separation between
\1h\ and each jet as a function of time.  In this figure, the origin of
time (day 0) is the day (2003 April 8 = MJD 52738) in which a major
radio flare, and hence a massive plasma ejection, was observed (Rupen
et al. 2003b). 

This figure clearly illustrates that the eastern jet is moving away
from \1h.  The {\em Chandra} and ATCA positions of the eastern jet are
consistent with a linear increase of the angular separation versus time
and a linear fit to the angular separation implies a proper motion of
15.2 $\pm$ 1.8 mas day$^{-1}$. The {\em Chandra} locations are
consistent with a continuation of the motion from the ATCA
observations. The data for the western jet are only weakly inconsistent
with a fixed position. A linear fit to the angular separation for the
western jet gives a proper motion of  6.7 $\pm$ 5.2 mas day$^{-1}$.

Extrapolation of the linear fit for the eastern jet (for which we have
the best statistics)  implies zero separation on MJD 52652 $\pm$ 38
(i.e. around 2003  January 13). This date is well before the initial
detection of \1h\ by INTEGRAL on 2003 March 21 (MJD 52719;  Revnivtsev
et al. 2003). With the RXTE/ASM data, it is not possible to assess the 
X-ray activity of \1h\ around MJD 52652. However, observations of several 
black hole systems (Corbel et al. 2004; Fender et al. 2004) has revealed 
that the massive ejection events occurred later the during the outburst 
at a period consistent with the transition from the intermediate state 
to the steep power-law state. Again, the bright radio flare observed  
by Rupen et al. (2003b) on 2003 April 8 (MJD 52737) is consistent with this
interpretation, as illustrated by e.g. the hardness evolution of the
X-ray spectra (Capitanio et al. 2005).

For the rest of this paper, we therefore consider the time origin of
the jets as the day of the bright radio flare (day zero in Figure 8).
With the linear fit to the motion of the eastern jet, we predict an
angular  separation of 1.29 $\pm$ 0.55 \arcsec\ at the time of the
first bright radio flare. If the velocity of the eastern jet  was
constant, we should have obtained a zero separation on this day.
Whereas the deviation from zero is not highly significant (2.4 sigma or 98\% confidence
level), this could imply that the jets were intrinsically more
relativistic at the time of ejection and decelerated gradually later on.
This would be another similarity with the  X-ray jets of \xte\ (Corbel et
al. 2002). More frequent radio and/or X-ray observations would have
been necessary to better constrain this motion.

The proper motion of the eastern jet appears higher that of the western
jet. Also, the X-ray flux of the western jet continues to rise and the
eastern jet is decaying in both the radio and X-rays. The western jet
is not detected at radio frequency during the April and June 2004 ATCA
observations, despite a sensitivity similar to the February 2004
observation (Table 2).  This could suggest that the rise of emission
from the western jet has stopped and that the peak of its emission
occurred sometimes between February and April 2004, i.e. few months
after the peak observed for the eastern jet. This probably indicates
that the eastern jet is approaching and the western  jet is receding.

\subsubsection{Case of ballistic jets}

As there is no strong indication of a decelerating jet in \1h , we first consider
the kinematics of the jets under the assumption of an intrinsically symmetric 
ballistic ejection (e.g. see Fender et al. 1999 for the formalism).  
From the first detection of the jets (ATCA on MJD 52956 for the 
eastern jet and {\em Chandra} on MJD 53092 for the western jet), we 
estimate their average proper motions between ejection on 2003 
April 8 and their first detections. We deduced an average proper motion of  
21.2 $\pm$ 1.4 mas day$^{-1}$ for the eastern jet and 13.3 $\pm$ 0.6 mas 
day$^{-1}$ for the western jet. From this, we derive $\beta cos \theta 
= 0.23 \pm 0.05 $, with $\beta = v/c$ the true bulk velocity and 
$\theta$ the angle between the axis of the jet and the line of sight.
This immediately gives a maximum angle to the axis of the jets of 
$\theta_{max} \le 77 \pm 3$\degr\ and a minimum velocity of $\beta_{min}
\ge 0.23 \pm 0.05$. We can also infer a maximum distance to \1h\ of
10.4 $\pm$ 2.9 kpc. To date, the distance to \1h\ is basically unknown, 
however its location toward the Galactic bulge could possibly imply a Galactic
Center location, which would be consistent with the above upper limit.  
If we assume a source distance of 8 kpc (distance to the Galactic center), 
we derive an intrinsic velocity of the ejection of $\beta = 0.79$ and
an angle of $\theta = 73$\degr\ for the axis of the jets. 
At this distance the apparent velocity of the eastern jet ($\beta_{app} 
= v_{app}/c$) is 0.98 and for the western jet we deduce an apparent velocity  
($\beta_{rec } = v_{rec}/c$) of 0.61. This clearly indicates the presence of
a significantly relativistic outflow in \1h , almost a year after the ejection
event. 

\subsubsection{Case of decelerating jets}

If the jets have been gradually decelerated since the ejection event,
the  average proper motions can not be used to constrain the jets
kinematics.   In that case, we can still study their properties using
the 2004  measurements of the proper motions (15.2 $\pm$ 1.8 mas
day$^{-1}$ for the eastern jet and 6.7 $\pm$ 5.2 mas day$^{-1}$ for the
western jet). We  again assume a source distance of 8 kpc, keeping in
mind that this  is a major source of uncertainty in the derived
parameters. At this  distance the apparent velocity of the eastern jet
($\beta_{app} =  v_{app}/c$) is 0.71 $\pm$  0.08 and for the western
jet we deduce  an apparent velocity  ($\beta_{rec } = v_{rec}/c$) of
0.31 $\pm$ 0.24.   The true bulk velocity, $\beta = v/c $, is defined
as $\beta =  \beta_{app}/(\beta_{app} {\rm cos} \theta + {\rm sin}
\theta) $, with $\theta$ equal to the angle between the axis of the jet
and the  line of sight. This function has a minimum at $\theta_{min}$ =
tan$^{-1}$(1/$\beta_{app}$) = 54\degr\ with our measured  apparent
velocity for the approaching jet (again for a distance of 8 kpc). This
translates into a minimum true bulk velocity of  $\beta_{min}$ = 0.57
$\pm$ 0.05, still indicating the presence in 2004 of a significantly
relativistic flow in \1h.

\subsection{Flux evolution and emission mechanism}

\subsubsection{Radio emission: Spectra}

We define the radio flux density S$_\nu$ $\propto$ $\nu^\alpha$, with
$\alpha$ the radio spectral index and $\nu$ the frequency. We note a
difference in the properties of radio emission during the rise versus
decay and the decay phases. The radio spectra are consistent with being
optically thin (i.e. $\alpha \le 0 $) all the time. We note that the
spectra are very steep during the rise (see Table 2, $\vert \alpha
\vert~\ge$~1), whereas the single radio detection during the decay is
more typical of optically thin synchrotron emission ($\alpha$ = $-$0.49
$\pm$ 0.22). The detection of a significant level of polarization (when
the sensitivity allowed it) is consistent with the radio emission being
optically thin synchrotron emission. The broadband spectra (see below)
accurately constrains the spectral index and is consistent with a
classical synchrotron spectral index of $\alpha \sim -0.6$.  This
index from the broadband SED on 2004 February 12-13  is significantly
different from the average radio spectral index during  the rise, which
is $\alpha$  = $-$1.18 $\pm$ 0.13.

The difference of spectral indices during the rise is probably not
related  to the emission beeing resolved out at high frequency by the
ATCA, as the  four radio observations during the rise were performed in
various configurations of the telescope. In addition, we note that both
jets are not resolved by  {\em Chandra}, and therefore they should
appear as point sources to ATCA.  Whereas the synchrotron slope is
almost a textbook example during the decay, the spectral index during
the rise is likely an intrinsic property of the process that
accelerates the emitting particles during the collision  of
relativistic plasma with the local medium. 
This may possibly be explained as followed: The rising phase represents the
propagation of the jets through a dense medium associated with a
mechanism that accelerate particles to high energy, whereas the
decaying  phase represents propagation through a less dense medium and
possibly  without any further re-acceleration of particles in the jets.
The emission  properties during the decay phase would then be governed
by synchrotron  and/or adiabatic losses of the relativistic particles.
The steep spectrum during the rise may eventually shed light on the
mechanism that accelerates particles to high energies.

\subsubsection{Broadband spectra and emission mechanism}

The eastern and western jets are simultaneously detected at radio and
X-ray frequencies on 2004 February 12-13.  Figure 9 shows the broadband
spectral energy distribution of the jets at this epoch. The radio
fluxes are from Table 2, and we convert the unabsorbed 0.3--8 keV flux
to a flux density at a pivot X-ray energy defined by allowing the X-ray
photon index to vary (Fig. 9).  
A combined fit of the radio and X-ray data result in
a spectral index of -0.64 $\pm$ 0.02, which is typical of synchrotron
emission.  The same is true for the western jet, giving a radio/X-ray
spectral index of -0.70 $\pm$ 0.05 on 2004 February 12-13.

The detection of linear polarization ($\sim$ 10\%) from the eastern
jets of \1h, as well as the spectral  index, strongly favor synchrotron
emission as the physical origin of the radio emission in the jets.  The
fact that the X-ray emission is consistent with an extrapolation of the
radio power-law suggests that the synchrotron emission extends up to
X-rays.  

We can therefore derive the minimum energy associated with the  eastern
jet on 2004 February 12-13 (the jet for which we have the best
constraints) under the assumption  of equipartition between the
magnetic and electron energy densities (Longair 1994).  We assume 
the case of ballistic ejection (section 3.1.1), corresponding to an angle 
between the jet axis and the line of sight of $\theta$ = 73\degr\ and
a bulk velocity $\beta$ = 0.79. We use a frequency range of 1.4 $\times 10^{9}$ Hz to 2
$\times 10^{18}$ Hz with a spectral index of --0.6  and a flux density
of 2.4 nJy at the pivot energy (Figure 9) of 2.7 keV. 

The major uncertainty for this calculation is the estimate of the
volume of the emitting region.  For that purpose, we use the same method as in 
Tomsick et al.\ (2003a) by assuming that the emitting region is a section
of a cone with its vertex at the compact object.  An upper limit can be
obtained using the {\em Chandra} constraint on the source size (FWHM
$\le$ 1.4\arcsec ) as calculated in section \ref{xmorph}.  The opening
angle of the cone is not well constrained, so we consider the upper
limit from the {\em Chandra} observations of 12\degr\ and also 1\degr\
as in \xte\ (Kaaret et al. 2003), giving a volume of the emitting 
region of about 4 $\times 10^{51}$ cm$^3$ (for an opening angle of
12\degr) or 3 $\times 10^{49}$ cm$^3$ (for 1\degr).   

For this volume range, the corresponding minimum energy required is
about 1.6 $\times 10^{42}$ to 1.4 $\times 10^{43}$ erg and the associated
magnetic field for which the energy in relativistic particles equals 
the magnetic energy is in the range 0.2 to 0.8 mG.  The kinetic energy
of a pure electron/positron plasma is $10^{42}$ to 9 $\times 10^{42}$ erg.  If
there is one proton per electron, then the kinetic energy of the
electron/proton plasma is about 2 $\times 10^{42}$ to  1.1 $\times
10^{43}$ erg.  The associated mass of this electron/proton plasma would
be of the order of 1$0^{21}$ to 4 $\times 10^{21}$ g. For a
typical mass  accretion rate of 10$^{18}$ g s$^{-1}$, this amount of
material could be accumulated in 1000 to 4000 s. As the mass outflow 
rate is likely much lower, an accumulation time of the order of a
day would be obtained if we assume a few percents efficiency, this could be 
consistent with the timescale of the initial radio flare. 

The Lorentz factor of the X-ray synchrotron emitting electrons would then be 
of the order of 1 to 3 $\times 10^7$ (taking a pivot energy of 2.7 keV and assuming
equipartition), giving synchrotron lifetime cooling of the orders of 2
to 50 years, i.e. much longer than the lifetime of the X-rays jets of
\1h.  It is interesting to note (keeping in mind the limitation on the
volume estimate) that all the above derived quantities are of the same
order of magnitude as those derived for \xte\ (Corbel et al. 2002;
Tomsick et al. 2003a; Kaaret et al. 2003). 

In the above, the X-ray emission seems consistent with an extrapolation
of the synchrotron radio spectrum.  However, it is of some interest to
see if some other emission mechanism, such as inverse Compton or synchrotron
self-Compton, could also be viable at high energy.  Assuming that only
the radio emission is of synchrotron origin lead to an equipartition
magnetic field of 0.2 to 0.7 mGauss; from this we deduce a range for
the magnetic energy density  of the order of 2 $\times 10^{-9}$ to 2
$\times 10^{-8}$ erg cm$^{-3}$. The synchrotron photon energy density 
is then estimated, similarly to the \xte\ case (Tomsick et al. 2003a), 
to be of the order of 6 $\times 10^{-17}$  to 2 $\times 10^{-15}$ erg
cm$^{-3}$.  The synchrotron photon density is well below the magnetic
energy density, implying that SSC is very likely not important at X-ray
frequencies.  As argued in Tomsick et al.\ (2003a), inverse-Compton
emission from the interstellar radiation field and the cosmic microwave
background are unlikely because the associated energy density is well
below the magnetic energy density.  All of these considerations point
to a coherent picture which is that the X-ray emission associated with
the jets of \1h\ is synchrotron emission and therefore very high energy
($>$ 10 TeV) particles are produced in those jets.

\subsection{A comparison with other X-ray jets}

With this discovery of transient X-ray jets in \1h, we can now come  to
the idea that such events may be a common occurrence associated with 
any X-ray binary in outburst. Indeed, after the original discovery of 
X-ray jets in \xte\ (Corbel et al. 2002), a relativistic ejection in
\gx\ has also been observed to later develop into a large scale jet,
possibly related to the interaction with the ISM (Gallo et al. 2004).
However, the western large scale jet of \gx\ has only been observed at radio 
frequencies (E. Gallo, private communication), possibly due its fast
decay rate.  Indeed, taking the radio flux density quoted in Gallo et 
al. (2004), we can estimate the time-scale associated to the two knots
in the large scale jets of \gx\ assuming an exponential or a power-law 
decay. The 1/{\it e}-folding times of the exponential decays are 47.3
$\pm$ 8.3 days and 57.0 $\pm$ 14.7 days, whereas the indeces of the
power-law decays are 5.9 $\pm$ 1.0 and 4.9 $\pm$ 1.0, respectively for
knot A and knot B.  These timescales are consistent with those of \1h,
but much faster than those of \xte\ (Kaaret et al. 2003). 
The X-ray observation of \gx\ was 6 months after the last detection
of the western large scale jet at radio frequencies (Gallo et al. 2004)
and any X-ray emission likely decayed away before the observation.

It is unclear why the decay could be faster in some  sources, but this
may be related to local ISM conditions. Also, as outlined by Wang et al. (2003),
the (already) fast decay of the X-ray emission from the eastern X-ray
jet in \xte\ was not consistent with a forward shock propagating through
the ISM. On the contrary,  if
the emission was driven by a reverse shock following the interaction
of the ejecta with the ISM, then the rapidly fading X-ray emission could be
the synchrotron emission from adiabatically expanding ejecta.
With a "much" faster decay in \1h\ or even \gx, it may also be possible
that the emission in these two cases is also related to  a reverse
shock moving back into the ejecta. In that case, it would be interested
to see if a reverse shock model (as in Wang et al. 2003)  could also
reproduce the flux and spectral evolution that we observed during the
rise of emission of the jets of \1h. Also the fast transition between
the rising and fading phases should also be explained.

In addition, we would like to mention the strong similarities with the 
neutron star system Sco~X--1, which shows the ballistic motion of two 
radio lobes located on each side of the compact star (Fomalont et al.
2001a,b). However, the Sco~X--1 lobes are powered by a continuous and
unseen ultra relativistic beam of energy (Fomalont et al. 2001a,b), 
possibly related to the fact that Sco~X--1 is always close to the 
Eddington luminosity for a neutron star. Similarly, as discussed
before, the bright radio flare in \1h\ on 2003 April 8 took
place around the transition to a steep power-law state, i.e. the
state which is believed to be close to the Eddington luminosity.
Otherwise, \1h, like most X-ray transients, is usually believed to be in
a state of low accretion rate (quiescence), and even if a large fraction of the 
accretion energy could be carried out by an outflow (compact jet)
in this accretion regime (Fender 2004), the total amount 
of energy would still be much less than in the brighter X-ray state.
Therefore, as we do not have any indication for a continuous 
and persistent powerful beam of energy in \1h, we believe that 
the evolution of the large scale jets in \1h\ and \xte\ is the result of 
one (or several in a short period) inpulsive 
ejection event(s) which later interact with the ISM.
 
As we have calculated above, many of the parameters (magnetic field,
particle energy, etc.) derived for \1h\ are consistent with those
obtained  for \xte, and therefore strengthen the similarities between
these two  sources.  Here, we point to another similarity.  When first
detected, both jets in \1h\ were at the same angular distance  ($\sim$
4 arcsec = 0.16 pc for a distance of 8 kpc) from the black hole
(however, we note that the eastern jets may have been active some time
before our first ATCA observations), i.e. \  the plasma originally ejected
(probably  around 2003 April 8) by the black hole has traveled the same
distance  before brightening again, possibly due to the collision with
denser ISM.  Similarly, in \xte\ the jets were observed to brighten at
the same angular distance ($\sim$ 22 arcsec = 0.5 pc for a distance of
5.3 kpc) from the black hole.  This could indicate that these two black
holes lie in a cavity or that the jets propagate through an evacuated
channel pre-existing to the outburst (e.g.\ Heinz 2002) and that they
''turn on'' again when they  hit a denser ISM phase.  

Whereas either the ejection or the ISM was not symmetric or homegeneous in the 
case of \xte\ (Kaaret et al.\ 2003), the evolution of the radio emission of both jets 
may also suggest a similar case in \1h. Indeed, on 2004 February 13, the 
western jet was at an angular separation of 4.47 $\pm$ 0.30 \arcsec\ and with
the kinematics of the jets as outlined previously, it should have passed 
through the 4.63 to 5.25 arcsecond region during the last two radio observations.
If the ejection was symmetric and the ISM homegeneous, the western jet should have 
been brighter at radio frequencies in April-June than in February 2004. 
The non-detection of the western jet in the last two ATCA observation may therefore
suggest a hint of asymmetry in the ISM density or in the ejection. However, a more
detailled lightcurve would have been necessary to confirm this.

\acknowledgments

The Australia Telescope is funded by the Commonwealth of Australia for
operation as a national Facility managed by CSIRO. {\em RXTE}/ASM
results are provided by XTE/ASM team at MIT.  This publication makes
use of data products from the Two Micron All Sky Survey, which is a
joint project of the University of Massachusetts and the Infrared
Processing and Analysis Center/California Institute of Technology,
funded by the National Aeronautics and Space Administration and the
National Science Foundation. We thank the referee for constructive
comments. SC acknowledges useful discussions with
S.\ Heinz.  PK acknowledges partial support from NASA Chandra grant
GO4-5039X. JAT acknowledges partial support from NASA Chandra grant 
GO3-4040X.

\begin{figure} \plotone{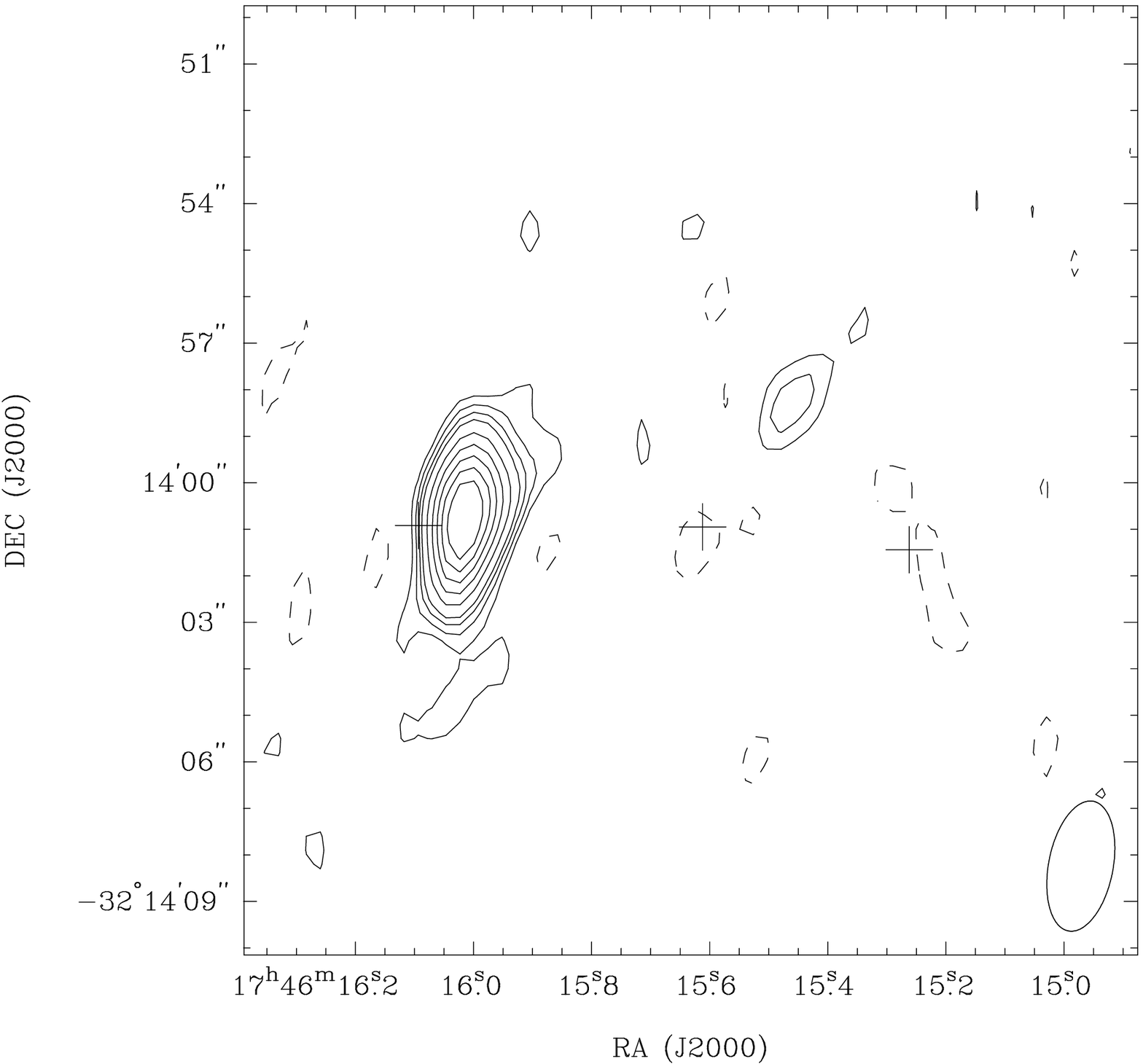} \caption{ATCA radio map at
8.64 GHz of the field near the black hole candidate \1h\ on 20 December
2003.  The crosses (size of 1\arcsec) indicate the location of \1h\ (center),  the
eastern and western jets, as observed during the first {\em Chandra} observations two months later
(on 12 February 2004). Contours are  plotted at -2, 2, 3, 4, 5, 7, 9,  11, 13, 15,
18, 21, 25 and 30 times the r.m.s. noise level of 0.07 mJy beam$^{-1}$. The
synthesized beam (in the lower right corner) is 2.8 $\times$ 1.4 arcsec$^2$, 
with the major axis at a position angle of --10.9 \degr.} \end{figure}

\begin{figure} \plotone{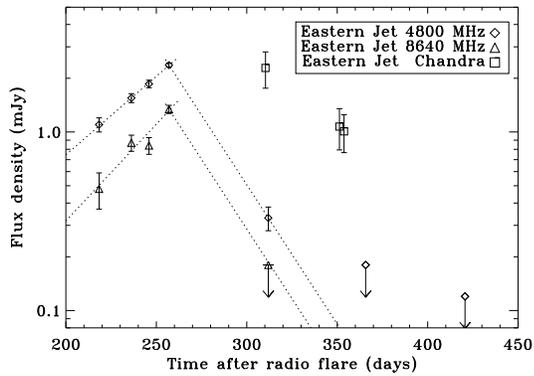} \caption{Radio light-curve at 8.6
GHz (3 cm) and 4.8 GHz (6 cm) of the eastern jet of \1h\ as measured by
ATCA. Upper limits are plotted at the three sigma level. The dotted
lines illustrate the exponential fit to the rise and decay of radio
emission. The {\em Chandra} 0.3--8 keV unabsorbed flux (times 10$^6$) 
of the eastern jet is also plotted and this indicates the dates of the X-ray
observations. The x-axis is the time since the major radio flare as observed by the VLA 
on 2003 April 8 (MJD 52738) by Rupen et al. 2003b.} \end{figure}

\begin{figure} \plotone{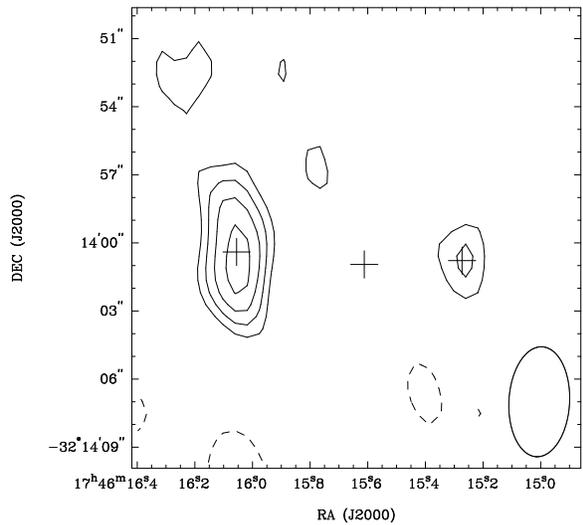} \caption{ATCA radio map at
4.8 GHz of the field near the black hole candidate \1h\ on 13 February
2004.  The crosses indicate the location of \1h\ (center),  the eastern
and western jets. The weak source at the position  of the western jet
is consistent with location the X-ray counterpart. Contours are 
plotted at -2, 2, 3, 4, 5 times the r.m.s. noise level of 0.05 mJy
beam$^{-1}$. The synthesized beam (in the lower right corner) is 4.8
$\times$ 2.7 arcsec$^2$, with the major axis at a position angle of --3.0 \degr.}
\end{figure}

\begin{figure} \includegraphics[height=19cm]{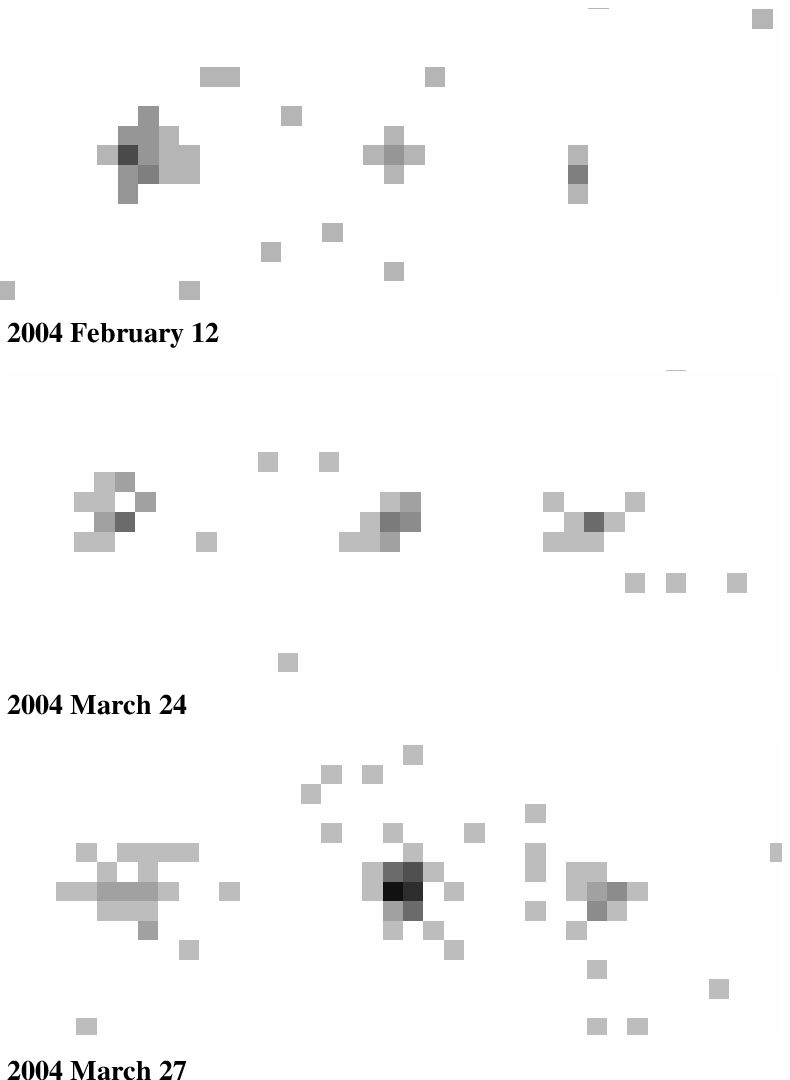} \caption{X-ray images of \1h\ for
the 0.3--8 keV band taken on 2004 February 12, March 24 and March 27.
The grey level represents the  number of X-ray counts per pixel with a
maximal of respectively 6, 5, 13 counts on resp. 2004 February
12, March 24 and March 27. \1h\ is located at the center of the image,
whereas the eastern jet is on the left and the western jet is on the
right.} \end{figure}

\clearpage

\begin{figure} \includegraphics[height=19cm]{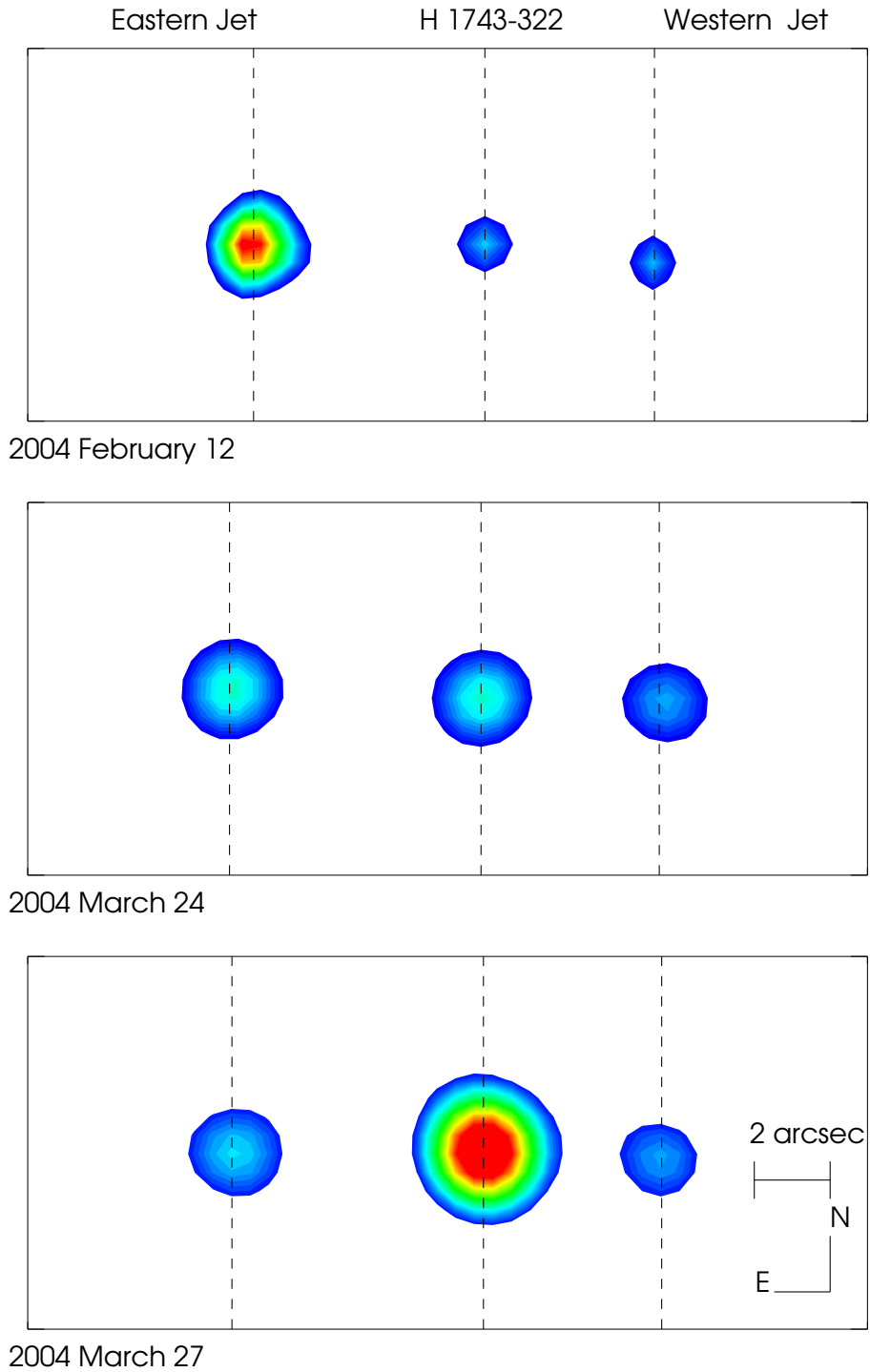} \caption{Filled contour plots
produced by convolving the 0.3--8 keV images shown in Fig. 4 with a
two-dimensional Gaussian  with a width of two pixels in both
directions. The vertical lines indicate the position of the X-ray sources 
in each observation. Each count image has been normalized by its integration
time.}
\end{figure}

\clearpage

\begin{figure} \plotone{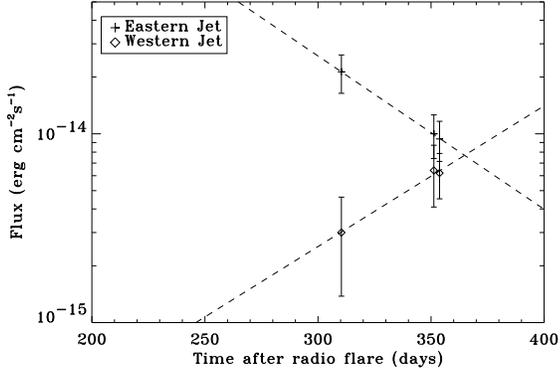} \caption{X-ray flux of the
eastern (plus) and western (diamond) jets versus the time after the
bright radio  flare on 2003 April 8. The curves are the exponential
decay (eastern jet) and ``rise'' (western jet) described in the text. }
\end{figure}

\begin{figure} \plotone{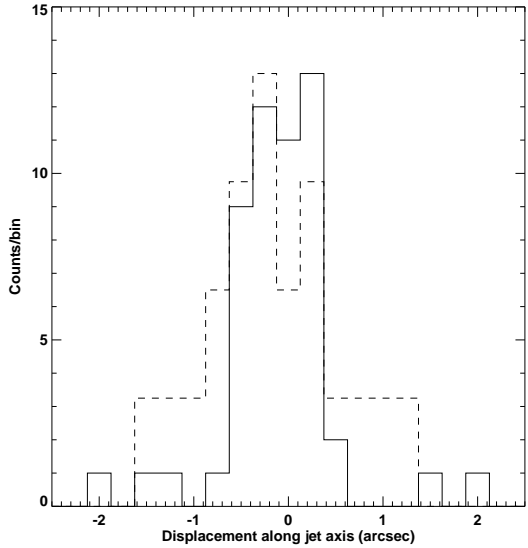} \caption{Distribution of the
X-ray counts along the jet axis combining {\em Chandra} observations \# 2 and
3. The dashed  line is the profile of the eastern jet and the solid
line is for \1h\ rescaled to match the peak of emission of the eastern
jet. The bin size is 0.25 \arcsec. } \end{figure}

\begin{figure} \plotone{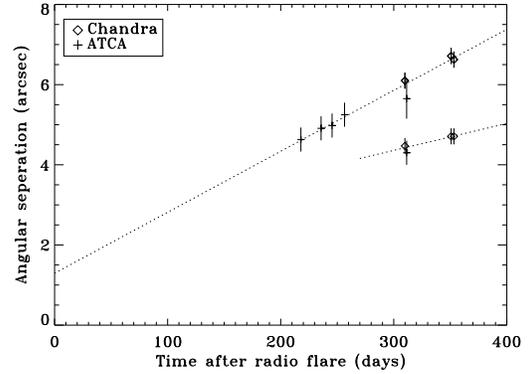} \caption{Angular separation
between the BHC \1h\ and each jet versus time since the bright radio
flare observed on 2003 April 8.  This plot is based on ATCA (plus sign)
and {\em Chandra} (diamond) data. The western jet is the one that is mowing 
slower.  The dotted lines represent the linear fit to the proper motion of the
jets assuming no deceleration. } \end{figure}

\begin{figure} \plotone{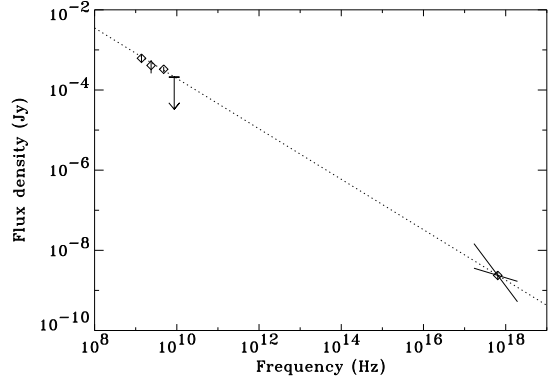} \caption{Spectral energy distribution
of the eastern jet on 2004 February 13 as observed by ATCA and {\em
Chandra}. The ''bow tie'' represents the {\em Chandra} constraints on
the flux and spectral index of the X-ray emission.  The spectral index
error lines are at 90\% confidence  level with the column density
frozen to value measured for the black hole \1h.} \end{figure}

\clearpage

\end{document}